# Symmetry-breaking and spin-blockage effects on carrier dynamics in single-layer tungsten diselenide

Ro-Ya Liu[1,2,3,4], Meng-Kai Lin[1], Peng Chen[5,6], Takeshi Suzuki[4], Pip C. J. Clark[7,†], Nathan K. Lewis[8], Cephise Cacho[9,10], Emma Springate[9], Chia-Seng Chang[2], Kozo Okazaki[4], Wendy Flavell[8,*], Iwao Matsuda[4,*], Tai-Chang Chiang[1,*]

[1] Department of Physics, University of Illinois at Urbana-Champaign, Urbana, Illinois 61801-3080, USA

[2] Institute of Physics, Academia Sinica, Taipei, 11529, Taiwan

[3] Advanced Light Source, Lawrence Berkeley National Laboratory, Berkeley, CA 94720, USA

[4] Institute for Solid State Physics, The University of Tokyo, Chiba, 277-8581, Japan

[5] Shanghai Center for Complex Physics, School of Physics and Astronomy, Shanghai Jiao Tong University, Shanghai 200240, China

[6] Key Laboratory of Artificial Structures and Quantum Control (Ministry of Education), Shenyang National Laboratory for Materials Science, School of Physics and Astronomy, Shanghai Jiao Ton University, Shanghai 200240, China

[7] Institute for Solar Fuels, Helmholtz-Zentrum Berlin für Materialien und Energie GmbH, Hahn-Meitner Platz 1D-14109 Berlin, Germany

[8] School of Physics and Astronomy and the Photon Science Institute, University of Manchester, Oxford Road, Manchester M13 9PL, United Kingdom

[9] Central Laser Facility, STFC Rutherford Appleton Laboratory, Harwell, OX11 0QX, United Kingdom

[10] Diamond Light Source, Harwell Campus, Didcot OX11 0DE, United Kingdom

* Corresponding authors: wendy.flavell@manchester.ac.uk; imatsuda@issp.u-tokyo.ac.jp; tcchiang@illinois.edu.

†Present address: Institute for Solar Fuels Helmholtz-Zentrum Berlin für Materialien und Energie GmbH Hahn-Meitner Platz 1, D-14109 Berlin, Germany; pip.clark@helmholtz-berlin.de

3# ABSTRACT

Understanding carrier creation and evolution in materials initiated by pulsed optical excitation is central to developing ultrafast optoelectronics. We demonstrate herein that the dynamic response of a system can be drastically modified when its physical dimension is reduced to the atomic scale, the ultimate limit of device miniaturization. A comparative study of single-layer (SL) tungsten diselenide ($WSe_2$) relative to bulk $WSe_2$ shows substantial differences in the transient response as measured by time- and angle-resolved photoemission spectroscopy (TRARPES). The conduction band minimum in bulk $WSe_2$, populated by optical pumping, decays promptly. The corresponding decay for SL $WSe_2$ is much slower and exhibits two time constants. The results indicate the presence of two distinct decay channels in the SL that are correlated with the breaking of space inversion symmetry in the two-dimensional limit. This symmetry breaking lifts the spin degeneracy of the bands, which in turn causes the blockage of decay for one spin channel. The stark contrast between the single layer and the bulk illustrates the basic carrier scattering processes operating at different time scales that can be substantially modified by dimensional and symmetry-reduction effects.

# I. INTRODUCTION

WSe$_2$ has attracted much attention for its remarkable mechanical, electronic, and optical properties [1] [2]. Its semiconducting band structure is well suited for optoelectronics applications. The bulk crystal structure in the 2H form is made of van-der-Waals-bonded Se-W-Se molecular layers [3]. Single- and few-layer WSe$_2$ can be readily prepared by exfoliation [4] or by growth via molecular beam epitaxy (MBE) [5]. Such ultrathin films are of interest for low-power high-packing-density electronics, and for potential engineering enhancement of their physical properties in the nanoscale or two-dimensional limit [4] [6].

Figure 1(a) shows the band structure of bulk WSe$_2$. The global conduction band minimum is located about midway between $\Gamma$ and K, which yields an indirect band gap of about 1.3 eV [7] [8]; this indirect gap size is very similar to that of Si. Because of the quasi-two-dimensional nature of the bulk lattice, illustrated in Fig. 1(c), the band dispersion along the layer normal is weak [7] [8] [9]. As expected, the band structure of SL WSe$_2$ is overall similar to that for the bulk, as illustrated in Fig. 1(b); yet there are important differences. Space inversion symmetry, present for the bulk crystal [10], is lacking in the SL, as evident from an inspection of its structure illustrated in Fig. 1(d). The spin-degenerate bands in the bulk are separated in the SL by spin-orbit-coupling. The conduction band minimum and the valence band maximum for the SL are both located at the $\overline{K}$ point and form a direct gap of approximately 1.3 eV [11]. Direct optical excitation across this gap using polarized light can result in a tunable spin texture [8] [12]. This spin-valley coupling effect offers an extra degree of freedom or control for optoelectronic information processing [2] [13] [14]. SL WSe$_2$ is also of interest for its quantum spin-Hall electronic structure when it is prepared in a metastable 1T' form [15].

This paper reports a pump-probe study using TRARPES. While some aspects of the dynamical behavior of this and related materials have been explored using a variety of techniques [16] [17] [18] [19], our work reported herein offers the first detailed side-by-side comparison of bulk and SL WSe$_2$ in regard to their ultrafast dynamic responses in the femtosecond (fs) to the picosecond (ps) time scales. The surprisingly large differences found in our experiment reveal a novel symmetry effect, which leads to spin-selective decay of the carriers.

## II. METHODS

### A. Sample synthesis and characterization

Experiments of SL WSe$_2$ growth, bulk WSe$_2$ cleavage, and *in situ* ARPES measurements were performed at the Advanced Light Source (ALS), Lawrence Berkeley National Laboratory. Wafers of 6H-SiC(0001) were outgassed at 680˚C for hours and processed by 80 cycles of flash-annealing to 1300˚C. The resulting C-face was a reconstructed SiC surface covered by a bilayer graphene (BLG) [20]. SL WSe$_2$ was grown on top with the substrate maintained at 400˚C by MBE using an electron-beam evaporator and a Knudsen cell for W and Se evaporation, respectively [15]; the resulting sample structure is schematically illustrated in Fig. 1(d). The growth process was monitored by ARPES and reflection high energy electron diffraction.

Figure 2(a) shows an ARPES spectrum for a SL sample taken with 42 eV photons to cover the entire range of $\overline{\Gamma K}$ and beyond. The observed band structure agrees fairly well with the theoretical band diagram in Fig. 1(b) with no features that can be attributed to two layers or thicker films, thus confirming the SL configuration [21] [22]. The additional features in Fig. 2(a) come from photoemission from the π bands of the underlying BLG, as labeled in the figure. These bands are Dirac-like and cross at 1.70 Å$^{-1}$, or the $\overline{K}$ point of the graphene Brillouin zone, at an energy just slightly below the Fermi level. Bulk WSe$_2$ samples were obtained by cleavage *in situ*. For a side-by-side comparison, Figs. 3(a) and (b) show ARPES maps along the $\overline{\Gamma K}$ direction for bulk and SL WSe$_2$, respectively, both obtained using 50 eV photon energy. Each map shows two valence bands with maxima at $\overline{K}$ or 1.27 Å$^{-1}$ in good agreement with the theoretical band structures.

## B. TRARPES experimental setup

Our TRARPES studies were performed with 730-nm (1.7-eV) pump pulses and 25-eV probe pulses. Both the pump and probe pulses were generated from a 1 kHz Ti:sapphire amplified laser system with an output wavelength of 785 nm, a pulse energy of 12 mJ, and a pulse duration of 30 fs [23]. Part of the 785 nm beam was focused into a cell filled with Ar gas for high harmonic generation (HHG); the 15-th order output at 25 eV was selected for the probe beam. An optical parametric amplifier (OPA, HE-Topas from Light Conversion) followed by a frequency mixing stage and a second harmonic generator provided the pump beam at 730 nm or 585 nm for our experiments. Only results from the 730-nm pump are shown here, but additional results from 585-nm pump are consistent. The fluence of the pump beam was kept at ~0.9 mJ·cm$^{-2}$ per pulse. Both the pump and probe beams were p-polarized. The energy, angular, and time resolutions of TRARPES were 400 meV, 0.3˚, and 50 fs, respectively.

Samples of SL WSe$_2$, after characterization at the ALS, were capped with a thick layer of Se for protection before shipping to Artemis, Central Laser Facility, Rutherford Appleton Laboratory, for TRARPES measurements. The capping layer was desorbed *in situ* by heating to 400˚C to expose the WSe$_2$ SL. Figure 2(b) shows a low energy electron diffraction pattern from the SL sample. Spots in the red circle and the orange square are 1×1 diffraction points of BLG and WSe$_2$, respectively. The two lattices are incommensurate. The WSe$_2$ spots show an azimuthal mosaic spread, which is typical for incommensurate growth, but its average lattice is crystallographically oriented in parallel to the substrate lattice.

Figures 3(c) and (d) show, respectively, ARPES spectra measured by using the HHG beam from a bulk WSe$_2$ crystal cleaved *in situ* and a SL sample originally capped under a Se protective layer but decapped *in situ* just before the measurements. The results are similar to those shown in Figs. 3(a) and (b), respectively, but the spectral features are significantly broader. Because the average photon flux from HHG is substantially lower than that from the synchrotron source, a wider analyzer slit and thus a lower energy resolution was necessary for acquiring data with sufficient statistics, especially for the SL sample. The significantly lower energy resolution makes the HHG spectral features broader.

### C. Fitting functions for the time evolution

We use the convolution of multiple exponential functions and a Gaussian function to fit the observed TRARPES decay processes. The temporal response function is described by

$$I(t) = \begin{cases} 0, & t < t_0 \\ \sum_{i=1}^{N} A_i\, e^{-\frac{t-t_0}{\tau_i}}, & t \geq t_0 \end{cases}$$

where $t_0$ is time zero, and $A_i$, and $\tau_i$ are the amplitude and decay time constant for the *i*-th exponential function, respectively. The Gaussian function representing instrumental broadening $G(t)$ is given by

$$G(t) = \frac{1}{\sqrt{2\pi}\sigma} \exp\left(-\frac{t^2}{2\sigma^2}\right),$$

where $\sigma$ is the standard deviation.

The fitting function *f*(t) is given by the convolution

$I(t) \otimes G(t)$

$$= \sum_{i=1}^{2} \int_{t_0}^{\infty} A_i \exp\left[-\left(\frac{t'-t_0}{\tau_i}\right)\right] \times \frac{1}{\sqrt{2\pi}\sigma} \exp\left[-\frac{(t-t')^2}{2\sigma^2}\right] dt'$$

$$= \sum_{i=1}^{2} \frac{A_i}{2} \text{erfc}\left(\frac{t_0 - t + \frac{\sigma^2}{\tau_i}}{\sqrt{2}\sigma}\right) \times \exp\left(-\frac{t - t_0 - \frac{\sigma^2}{2\tau_i}}{\tau_i}\right)$$

### III. RESULTS AND DISCUSSION

#### A. Bulk WSe$_2$

Figure 4(a) shows a differential TRARPES map taken from bulk WSe$_2$, obtained by taking the difference between the ARPES map obtained at 26 fs delay after the arrival of the pump pulse and an ARPES map obtained just before the arrival of the pump pulse. The red and blue colors indicate positive and negative differences, respectively. The total time resolution of the experiment is about 50 fs. Since the 26 fs delay is less than the time resolution of the experiment, the difference map highlights the prompt response of the system. A prominent feature is a red spot at approximately 1.27 Å$^{-1}$ and 0.15 eV above the Fermi level, enclosed in a box, which corresponds to transient carriers excited into the local conduction band minimum at the K point of WSe$_2$.

Figure 4(c) shows the integrated intensity of the red spot over the box as a function of delay time. The transient behavior is a prompt rise at time zero followed by a very fast decay. It can be well described by a single exponential decay following the optical excitation. Fitting of the results yields a decay time of $\tau^{BC} = 47 \pm 4$ fs (BC stands for bulk conduction band minimum). Thus, the transient carriers at the K valley decay essentially instantaneously within our time resolution of 50 fs. This very fast decay can be attributed to intra-band, inter-valley scattering to states near the global conduction band minimum located between Γ and K [24] [25]. The process is illustrated in detail by the conduction band diagram in Fig. 5(a). The global conduction band minimum is at a lower energy than the local conduction band minimum at K, and there is considerable phase space available for the carriers at the K point to scatter into. Because the bulk conduction bands are spin degenerate, the decay process is the same for both spin channels (indicated by red and blue colors in Fig. 5(a).

The valence band region in Fig. 4(a) shows alternating red and blue bands, which are indicative of energy shifts and broadenings of the valence bands following optical excitation. This behavior is common for this type of pump-probe measurements [26]. Figure 6(a), a reproduction of Fig. 4(a) and included here for easy reference, shows a pattern of bands colored in, from top to bottom, red, blue, red, blue, and red. The results suggest that the valence bands broaden in energy after pumping, and thus there is excess intensity (red color) both above and below the valence band region. Furthermore, the top red band is much more intense than the bottom red band, indicating an upward shift of the bands. The color bands appear uniform over the entire probed $k$ range, suggesting that the broadening and shift do not depend significantly on the wave vector $k$. The details of the changes are illustrated by the experimental energy distributions curves (EDCs) at K in Fig. 6(b) taken before (black curve) and at 26 fs after pumping (magenta curve). The difference (teal curve) shows a peak at the local conduction band minimum that corresponds to photoexcited carriers discussed earlier. In the valence band region, it shows, for decreasing energy, positive, negative, positive, negative, and weakly positive differences.

A detailed analysis yields the energy broadening of the EDCs averaged over $k$ as a function of delay time. The transient behavior for the energy broadening, shown in Fig. 6(c), is characterized by an amplitude of $52 \pm 5$ meV for the initial jump and a decay time constant of 2.2

± 0.2 ps. The same analysis yields the energy shift as a function of delay time; the transient behavior, shown in Fig. 6(d), is characterized by a much smaller amplitude of 7 ± 3 meV and a decay time constant of 6 ± 4 ps. The orders of magnitude for these decay time constants (several picoseconds) are consistent with typical time scales for phonon and lattice excitation, relaxation, and dissipation effects [27] [28].

## B. SL WSe$_2$

A differential TRARPES map for the SL at 33 fs delay excited by the same 730 nm pump shows an emission feature at the local conduction band minimum at $\overline{K}$, as indicated by the box in Fig. 4(b). It looks similar to the bulk case. However, its delay-time dependence, shown in Fig. 4(d), is evidently slower and more complex. It is characterized by an initial rapid decay followed by a much slower decay reaching beyond 1 ps. The behavior is well described by two exponential decays, one fast component with time constant $\tau_1^{SC} = 100 \pm 50$ fs and a slow component with time constant $\tau_2^{SC} = 700 \pm 500$ fs (here, SC stands for single-layer conduction band minimum). The slow-decay component is indicated in Fig. 4(d) by green shading. The two components have similar intensities with $(A_1\tau_1)/(A_2\tau_2) = 0.6 \pm 0.3$. This two-component behavior can be understood in terms of the schematic conduction band diagram shown in Fig. 5(b). In contrast to the bulk case illustrated in Fig. 5(a), the bands for the SL are generally spin-split for the lack of space inversion symmetry. The two bottom conduction bands in Fig. 1(b) and 5(b) are non-degenerate and have opposite spin directions, as indicated by the red and blue color coding. Inter-valley scattering for the blue band conserves the electron spin and is allowed, as indicated pictorially in Fig. 5(b), but the same process is suppressed for the red band. As a result, two decay times are expected, one fast and one slow, which correspond to the decay processes associated with the blue and red bands, respectively. Experimentally, the fast decay time $\tau_1^{SC}$ for the single layer is longer than $\tau^{BC}$ for the bulk case. The slower decay rate for the SL can be attributed to the reduced phase space available for decay. As seen in Fig. 5, the conduction band minimum for the SL is not much lower in energy than the local minimum at $\overline{K}$. Decay for the red band, characterized by a much larger time constant $\tau_2^{SC}$, is possible through intra-band spin-flip transitions, which are forbidden to first order and therefore should be much slower [19].

The dynamic behavior of the valence bands of the SL has been similarly analyzed as the bulk case to yield the energy broadening [Fig. 7(a)] and energy shift [Fig. 7(b)] as a function of delay time after optical pumping. The energy broadening is characterized by an exponential decay with an amplitude of $25 \pm 3$ meV and a decay time constant of $500 \pm 100$ fs. The energy shift is characterized by an amplitude of $22 \pm 2$ meV and a decay time constant of $270 \pm 90$ fs. The faster decay rates for the SL relative to the bulk may be attributed to a strong Coulomb interaction in the SL resulting from quantum confinement of the excited carriers [29]. The confinement can lead to enhanced carrier-carrier scattering, and thus a faster decay. Similar and related observations with varied results have been reported for other SL systems including $MoS_2/SiC$ [30], $MoS_2/Au(111)$ [31], and $WS_2/Au(111)$ [12]. However, SLs prepared on metal substrates can behave quite differently because of coupling to the abundant electrons and holes in the metal substrates.

### C. Dirac bands in the bilayer graphene under the $WSe_2$ SL

For the SL sample, TRARPES signal from the underlying BLG is readily detected as seen in Fig. 3(b). The differential TRAPRES map at 33 fs delay time, presented in Fig. 8(a), shows an enhanced (reduced) population of the Dirac bands above (below) the Fermi level as revealed by the intensity changes. The integrated intensity within the box in Fig. 8(a) is given in Fig. 8(b) as a function of time; fitting yields a decay time constant of $70 \pm 10$ fs (Fig. 7b). A similar analysis for the intensity reduction below the Fermi level yields the same time constant. These decay times have about the same magnitude as our time resolution, and the fast recovery from photoexcitation can be attributed to relatively efficient intraband scattering processes. Our observations are similar to those for graphene [32]. It appears that interlayer coupling across the heterojunction does not substantially affect the carrier dynamics in the graphene layers.

## IV. CONCLUSIONS

The starkly different time dependencies for the decay of the optically excited conduction band carriers between bulk and SL $WSe_2$ revealed by our side-by-side comparative study demonstrate that dimensional effects can play a critical role in the design of ultrafast optoelectronics. Generally, intra-band processes, when allowed, lead to a very fast decay of carriers after photoexcitation. Indeed, our measurements of the transient conduction band population in bulk $WSe_2$ confirm that the decay time is within our time resolution of 50 fs. A key finding is that for SL $WSe_2$, where the bands are spin split for lack of inversion symmetry, two decay times are observed, one fast and one slow. A physical picture derived from our analysis is that the slow decay channel is associated with spin flip, which is forbidden to first order. This spin-blockage effect, potentially useful for spin control and processing, is of interest for spintronic applications at the ultrafast time scales. Our study also reveals the time dependencies of the valence bands and the Dirac bands of the graphene layers under the SL sample. The transient behavior, documented herein, provides a useful reference for further studies of other related materials; contribution to the data base is essential for establishing a general understanding of the behavior of carrier dynamics in transition metal dichalcogenides and other layered systems.

## ACKNOWLEDGEMENTS

This work is supported by the U.S. Department of Energy, Office of Science, Office of Basic Energy Sciences, Division of Materials Science and Engineering, under Grant No. DE-FG02-07ER46383 (TCC). The Advanced Light Source is supported by the Director, Office of Science, Office of Basic Energy Sciences, of the U.S. Department of Energy under Contract No. DE-AC02-05CH11231. IM, WF, and TCC organized the project. PC prepared the SL samples and performed ARPES at ALS. RYL, MKL, TS, PCJC, NKL, CC, ES, IM performed TRARPES measurements at Artemis, where CC and ES also provided technical help for the laser operation. RYL, MKL, and TCC analyzed the data, interpreted the results, and wrote the first draft. All coauthors contributed to discussions and improvements that led to the final manuscript. We thank A. Wyatt, A. Jones, R. Chapman and P. Rice for technical support during the Artemis beam time.

**Figure 1.** (a) Calculated band structure of bulk $WSe_2$. (b) Calculated band structure of SL $WSe_2$. For the conduction bands, red and blue colors indicate spin-split states, while green color indicates spin degeneracy. The BLG bands are indicated by orange curves. (c) A schematic side view of the atomic structure of bulk $WSe_2$. (d) A schematic side view of the atomic structure of the SL sample including the BLG-on-SiC substrate. Atoms of Se, W, and C are represented by gold, blue, and grey balls, respectively.

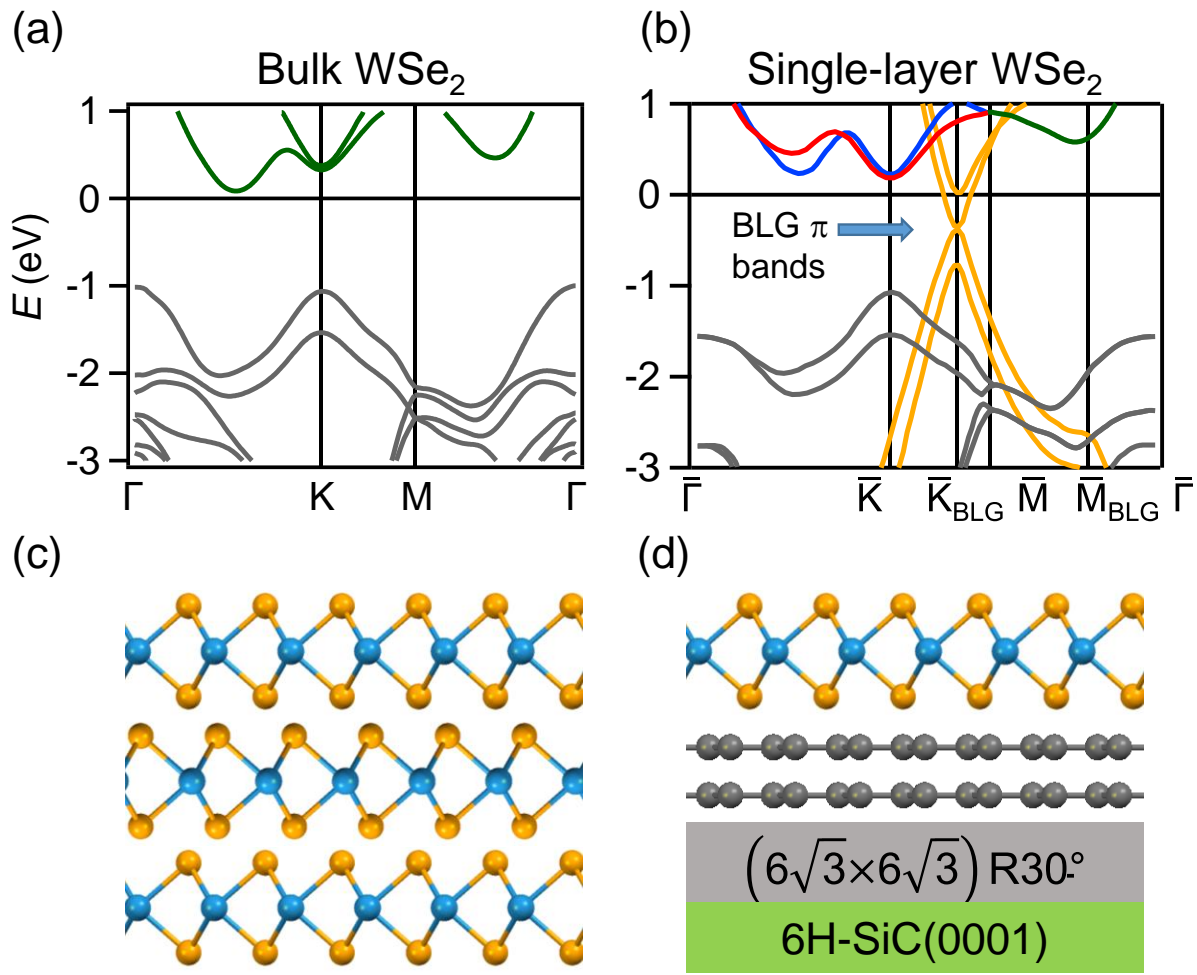

**Figure 2.** (a) ARPES map of a freshly prepared SL WSe$_2$ sample along $\overline{\Gamma}\overline{K}$ taken with 42-eV photons. The π bands from the underlying BLG are indicated. (b) Low energy electron diffraction pattern taken at 70-eV kinetic energy from a SL WSe$_2$ sample after decapping just prior to TRARPES measurements. Spots in the red circle and the orange square are 1×1 diffraction points of BLG and WSe$_2$, respectively.

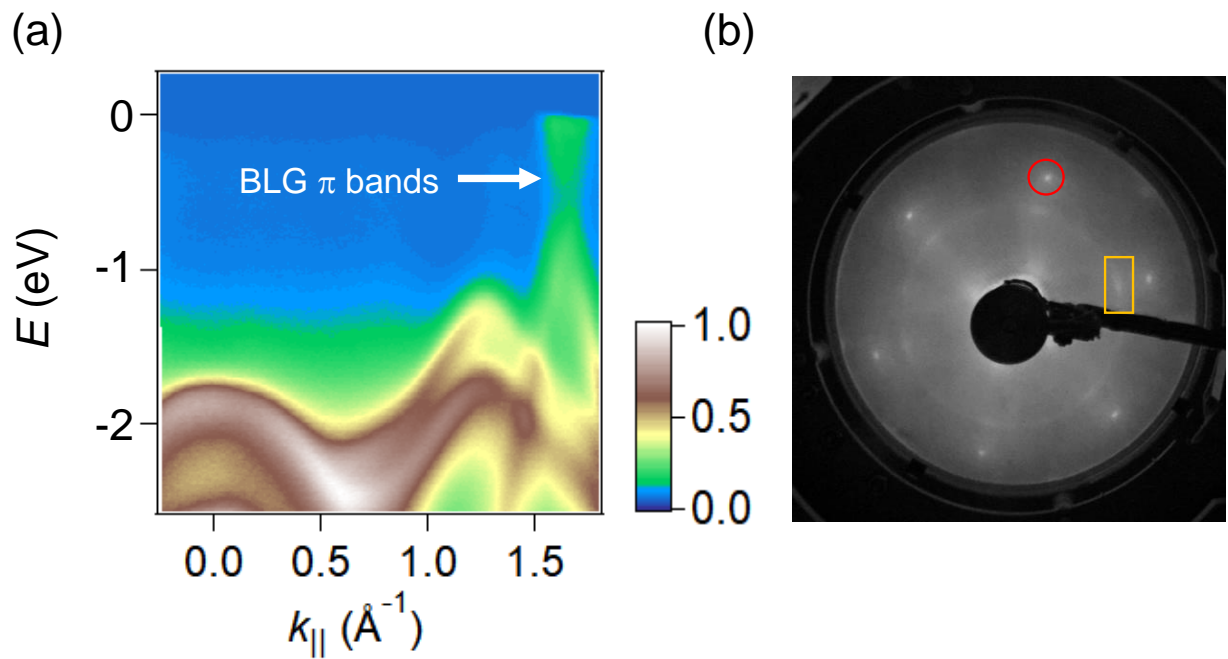

**Figure 3.** (a) ARPES map for bulk WSe$_2$ along $\bar{\Gamma}\bar{K}$ taken with 50 eV photons from a synchrotron. Two valence bands reach a maximum at $\bar{K}$ (1.27 Å$^{-1}$). (b) Similar ARPES map for SL WSe$_2$. The π bands from the underlying BLG are seen around $\bar{K}_{BLG}$ (1.70 Å$^{-1}$). (c) Similar ARPES map for bulk WSe$_2$ obtained using an HHG beam of 25 eV photons. (d) Corresponding results for the SL sample. The color-coded intensity scale is the same as that in Fig. 2(a).

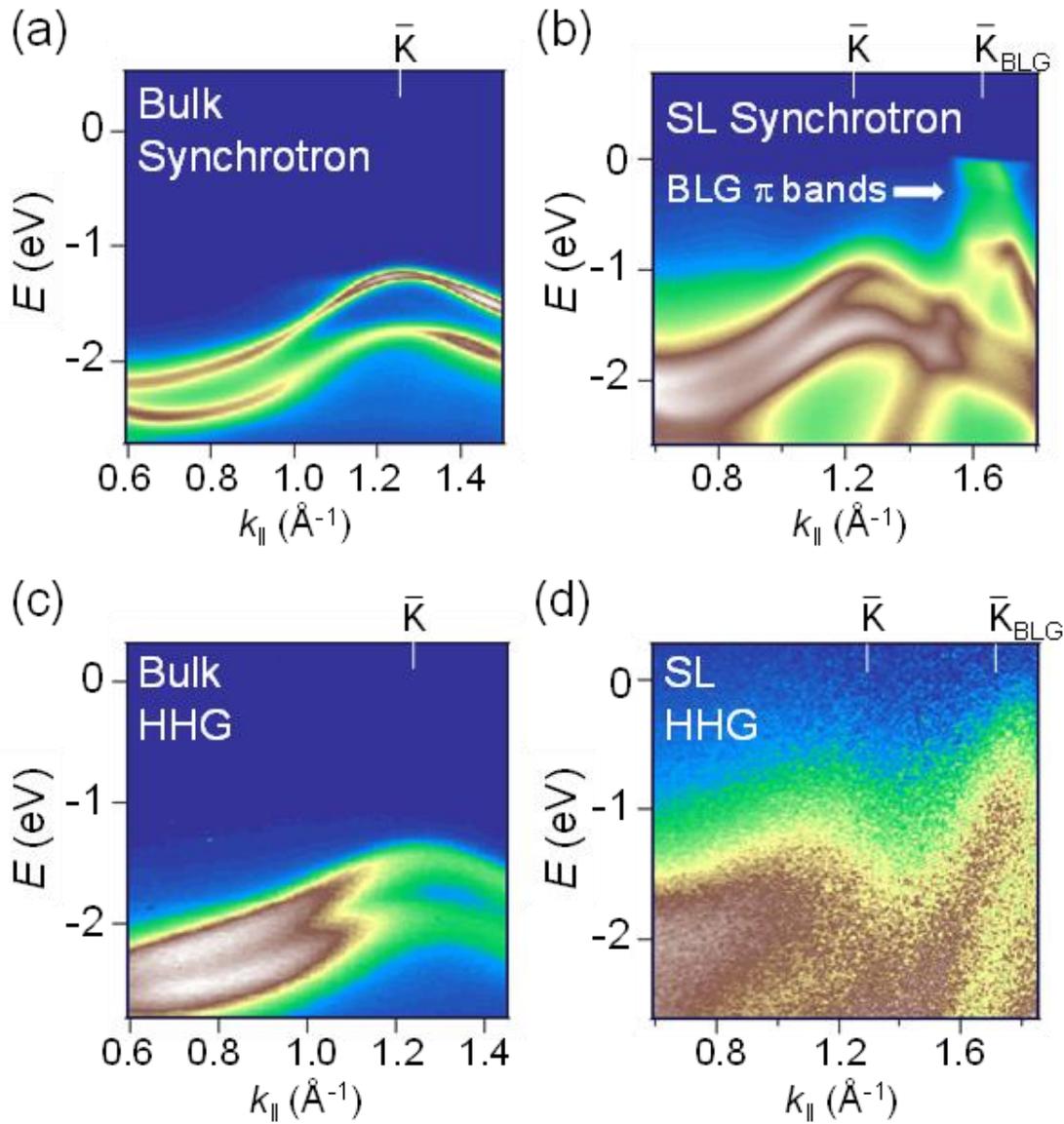

**Figure 4.** (a) Differential TRARPES map of bulk WSe$_2$ along $\overline{\Gamma}\overline{K}$ at a delay time of 26 fs. The red spot enclosed in a box is emission from carriers excited into the conduction band (CB) minimum at $\overline{K}$. (b) Similar ARPES map of SL WSe$_2$ at a delay time of 33 fs. (c) The integrated intensity within the box in (a) as a function of delay time. The behavior is well described by a single exponential decay. (d) Similar results for the SL. The decay of the intensity following excitation is well modeled by two exponential decays with very different decay times. The slow component is indicated by the green shading.

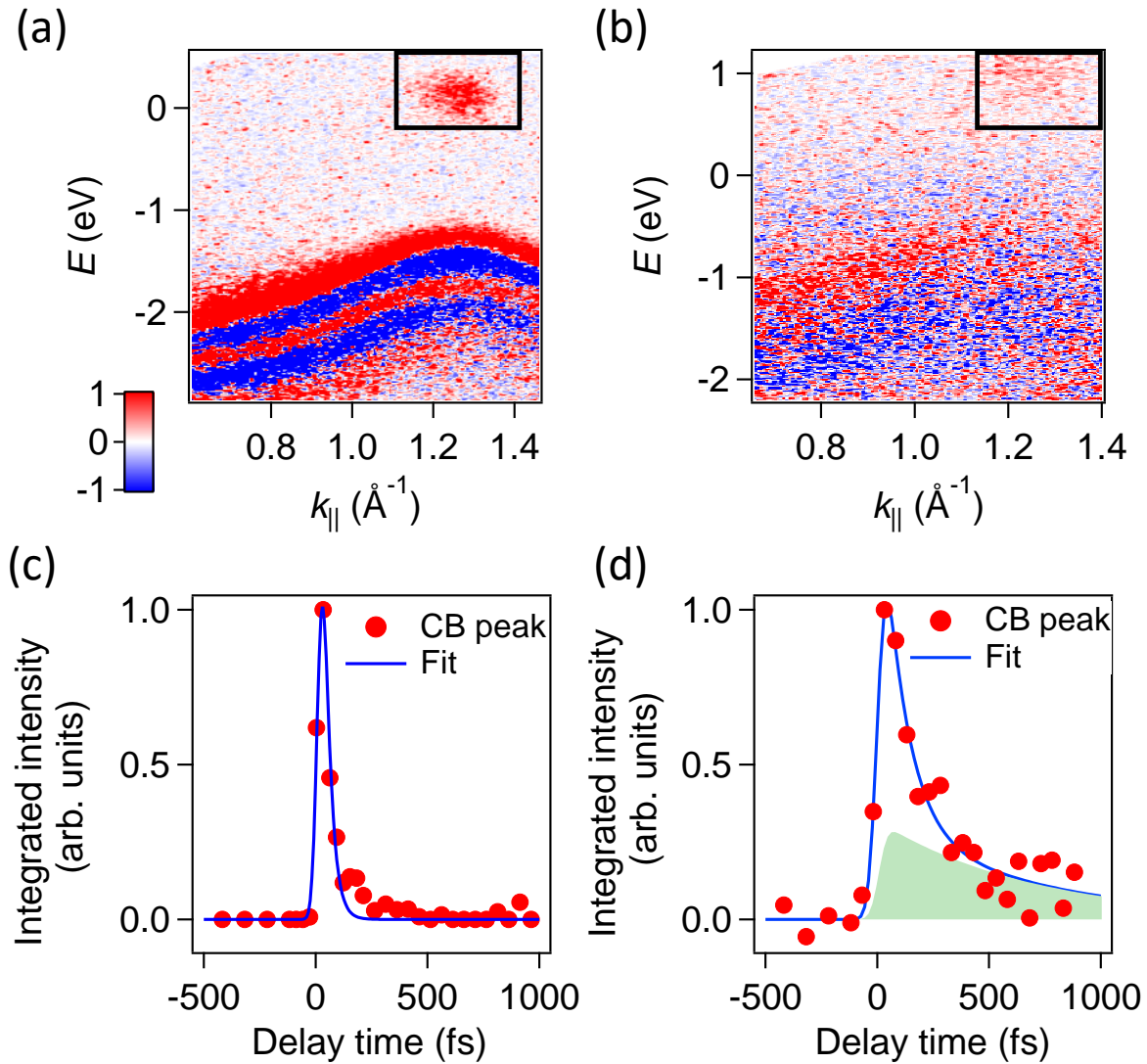

**Figure 5.** (a) Detailed view of the conduction band structure of bulk WSe$_2$. Each conduction band is spin degenerate. Carriers initially excited into the conduction band minimum at K decay by intra-band, inter-valley scattering to states near the global conduction band minimum located between Γ and K. The decay is allowed for both spin channels (red and blue), as indicated by the arrows. (b) A similar diagram for SL WSe$_2$. The bands are spin-polarized (red and blue) and non-degenerate. The blue carriers initially excited into the conduction band minimum at $\bar{\text{K}}$ can decay quickly with its spin preserved. The red carriers decay via slower spin-flip processes.

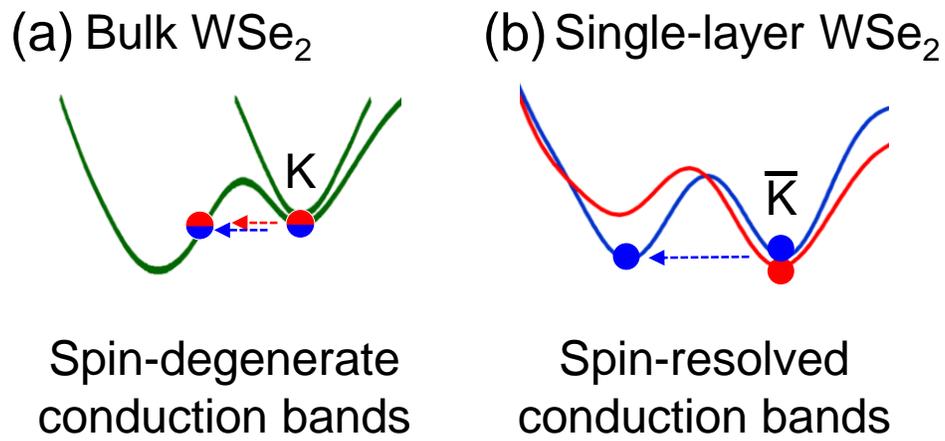

**Figure 6.** (a) Differential TRARPES map of bulk WSe$_2$ along $\overline{\Gamma}\overline{K}$ at a delay time of 26 fs. (b) Experimental EDCs at the zone corner $\overline{K}$ point taken before (black curve) and 26 fs after pumping (magenta curve). The teal curve represents the difference. (c) Energy broadening of the EDCs averaged over $k$ as a function of delay time. (d) Energy shift averaged over $k$ as a function of delay time.

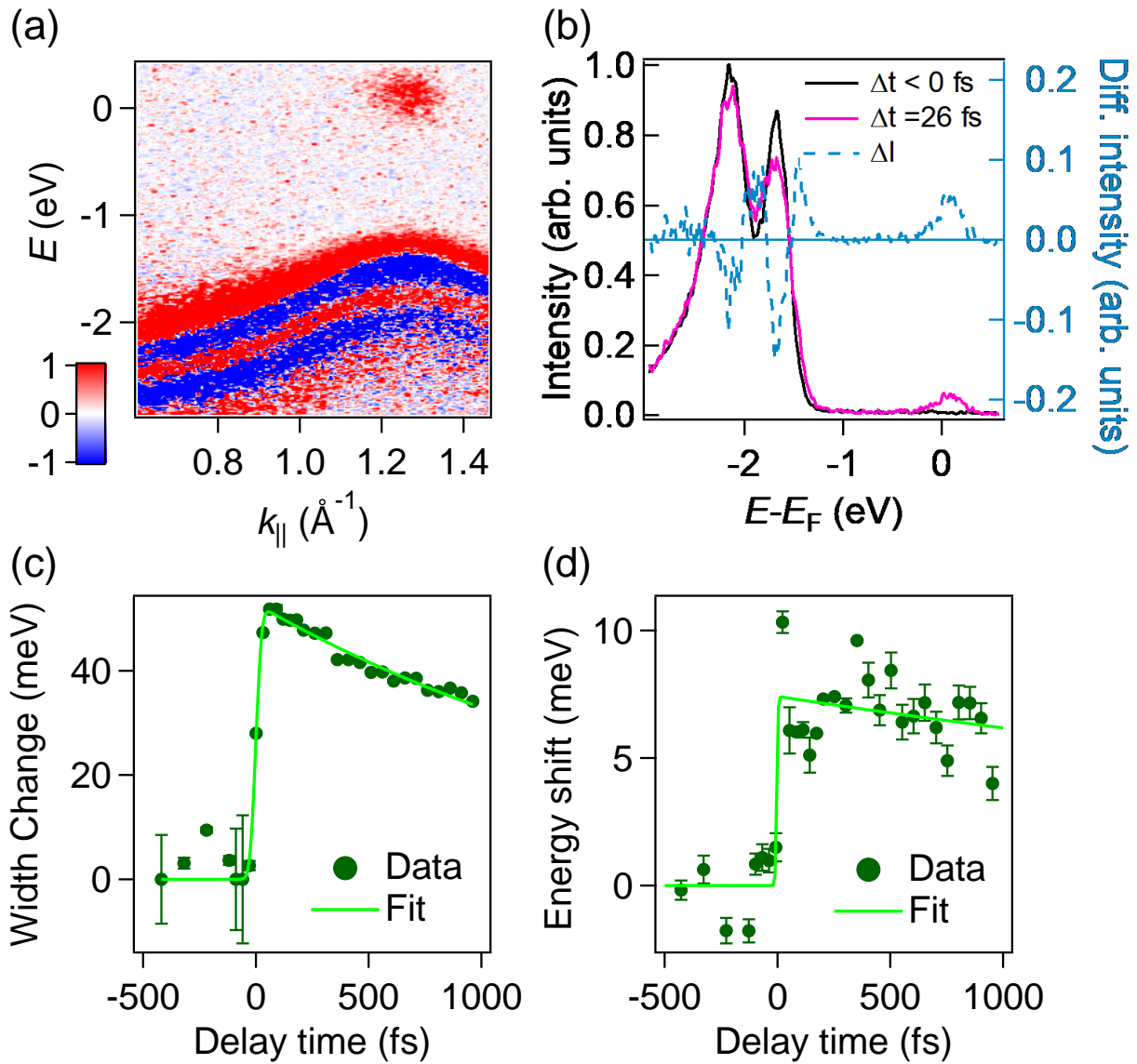

**Figure 7.** (a) Analysis of the TRARPES valence bands for SL WSe$_2$ yields the energy broadening as a function of delay time after optical pumping. The method of analysis is similar to that for the bulk case shown in Fig. 6(c). (b) Corresponding energy shift.

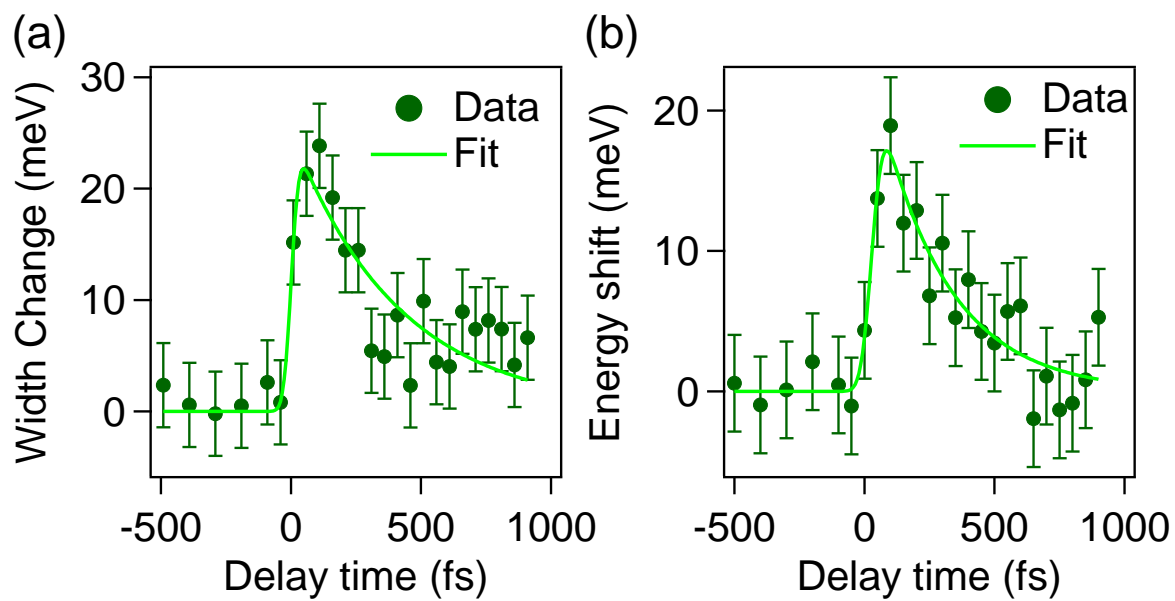

**Figure 8.** (a) TRARPES results at 33 fs delay from the BLG show an enhanced (reduced) population of the Dirac bands above (below) the Fermi level as revealed by the intensity changes. (b) The integrated intensity within the box in (a) as a function of delay time.

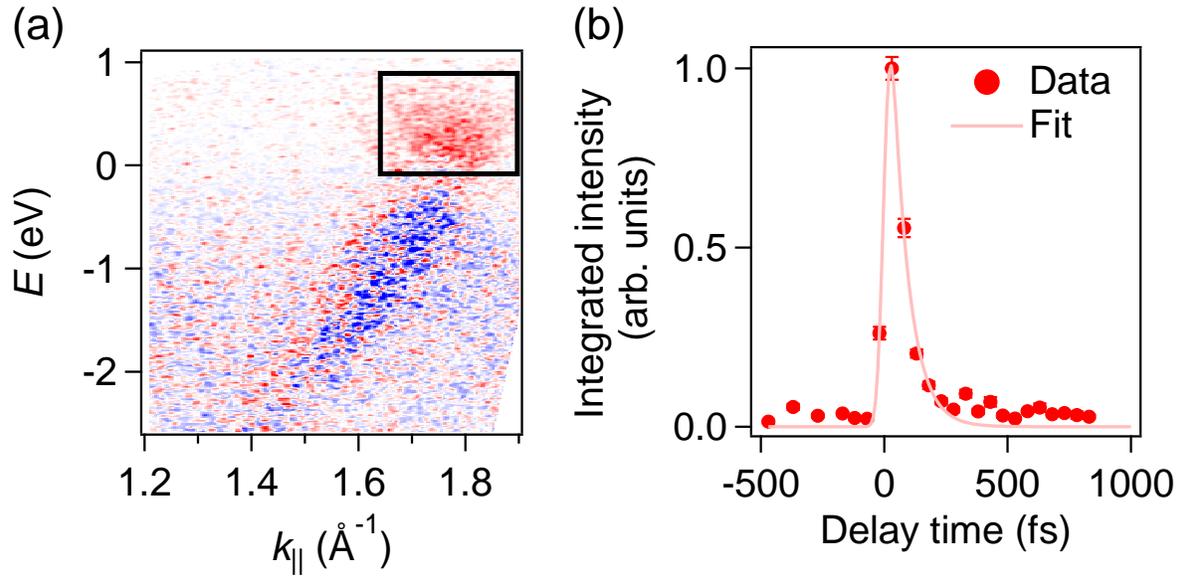